\documentclass[summary]{emts2019}
\usepackage{color}

\title{Applications and Potentials of Reciprocal Bianisotropic Metasurfaces}

\author{Mohammad Albooyeh$^*$\affref{ref1}, Viktar Asadchy\affref{ref2}, Jinwei Zeng\affref{ref1}, Hamidreza Kazemi\affref{ref1}, and Filippo Capolino\affref{ref1}}

\affiliation{%
  \aff{ref1}{Department of Electrical Engineering and Computer Science, University of California, Irvine, CA 92617, USA}
  \aff{ref2}{Department of Electronics and Nanoengineering, Aalto University, P.O. Box 15500, FI-00076 Aalto, Finland.}
  $^*$ mohammad.albooyeh@gmail.com
}

\begin{document}

\maketitle

\begin{abstract}
We discuss different applications and potentials of reciprocal bianisotropic metasurfaces including uniform and gradient metasurfaces, in particular, with anisotropic, chiral, and omega properties. Based on an analytic analysis, we discuss the necessary conditions to observe asymmetric co- and cross- polarization reflection and/or transmission, polarization conversion and rotation, asymmetric absorption, etc. We consider two kinds of incident wave scenarios based on linear and circular polarization.

\end{abstract}

\section{Introduction}

During the last decade, metasurfaces, i.e., conformal arrays of subwavelength resonant inclusions, have attracted enormous interest due to their  fascinating applications in the field of electromagnetic engineering. To name only a few, one may refer to negative index of refraction, perfect absorption of light, polarization rotation and transformation of the electromagnetic waves, subwavelength focusing of light, and in general arbitrary manipulation of the electromagnetic wave fronts. The essence of obtaining many of these alluring applications lies in the fact that one provides a ground to simultaneously take  advantage of the interaction between both components of electromagnetic fields (represented by the electric and magnetic field vectors ${\bf E}$ and ${\bf H}$, respectively) and electric and magnetic properties of materials (described by electric and magnetic susceptibility couplings $\chi_{\rm e}$ and $\chi_{\rm m}$, respectively). More applications, which are not otherwise achievable, including asymmetric reflection, refraction, and/or absorption for opposite illumination directions are obtained by exploiting cross-coupling magnetoelectric and electromagnetic susceptibilities $\chi_{\rm em}$  and $\chi_{\rm me}$.

Metasurfaces which exhibit a magnetic (electric) polarization current when excited by an electric (magnetic) field are named bianisotropic~\cite{biama}. As stated, such a behavior is modeled by cross-coupling magnetoelectric and electromagnetic parameters such as effective susceptibilities, polarizabilities, etc. From the reciprocity point of view, bianisotropic metasurfaces can be composed of both reciprocal and nonreciprocal inclusions (latter ones require an external  bias field, time-modulated or nonlinear materials). The material properties which define a general bianisotropic metasurface are four second-ranked tensors (dyadics).
There are various kinds of metasurfaces: metasurfaces that are composed of identical inclusions which are uniformly distributed as an array (uniform metasurfaces); metasurfaces that are composed of inclusions which are put together to create a gradient phase (gradient metasurfaces); or metasurfaces that are composed of different inclusions which are randomly distributed with nonuniform distances (amorphous metasurfcaes), they can be planar or conformal with different shapes such as cylindrical, spherical, etc. However, as long as the structure thickness and the arrayed inter-particle distances are subwavelength and the resonant behavior of the whole structure is directly defined by its inclusions' resonances and not by the inclusions' distances, it is referred to as a metasurface.

In the present study, we discuss some important observations about reciprocal, uniform or gradient bianisotropic metasurfaces by the use of a general analytical modeling. Moreover, we investigate different applications and potentials of metasurfaces which are only possible using bianisotropic chiral or omega properties. To that end, we present expressions for the transmission/reflection coefficients in term of the metasurface characteristic parameters, here, collective polarizabilities. Due to length limitation, we only discuss some important observations in this manuscript while we will have a complete discussion at the time of the presentation.

\section{Uniform Metasurfaces}
\begin{figure}[htbp]
  \centering
  \includegraphics[width=70mm]{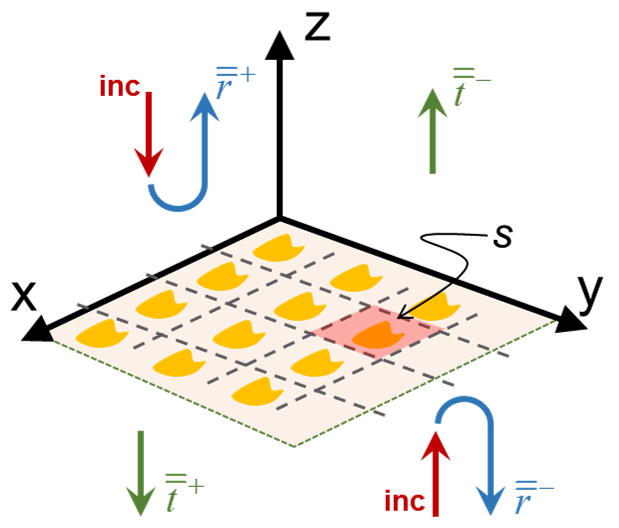}
  \caption{The geometry of a metasurface composed of identical inclusions which are uniformly distributed in a regular arrangement. The metasurface is either excited from top (towards $-z$ direction) or bottom (towards $+z$ direction), in each case the reflection and transmission tensors are denoted by $\bar{\bar{r}}^{\pm}$ and $\bar{\bar{t}}^{\pm}$, respectively.}
  \label{fig:fig1}
\end{figure}

Let us consider a homogenizable metasurface composed of similar inclusions which are uniformly distributed in a two-dimensional planar array. Without  loss of generality we assume that the metasurface is located in the $xy$-plane. For our discussion, we consider the case when the metasurface is illuminated normally by either a linearly polarized (LP) or a circularly polarized (CP) plane wave from either the froward (top) or backward (bottom) direction [see Fig.~\ref{fig:fig1}]. 

\subsection{Illumination with LP Plane Waves}
Let us start from the constitutive relations between the incident electric $ {\bf E}^\textrm{i}$ and magnetic $ {\bf H}^\textrm{i}$ fields and the induced electric $ {\bf P}$ and magnetic $ {\bf M}$ surface polarization densities of the metasurface which read~\cite{biama}\begin{eqnarray}
  \nonumber {{\bf   P}} &=& {\bar{\bar{\alpha}}^{\textrm{ee}}\over S} \cdot {\bf E}^\textrm{i}+{\bar{\bar{\alpha}}^{\textrm{em}}\over S} \cdot {\bf H}^\textrm{i}, \\
  {{\bf   M}} &=& {\bar{\bar{\alpha}}^{\textrm{me}}\over S} \cdot {\bf E}^\textrm{i}+{\bar{\bar{\alpha}}^{\textrm{mm}} \over S} \cdot {\bf H}^\textrm{i}, \label{our_collective}
\end{eqnarray}
respectively. Here $S$ is the unit cell area and $\bar{\bar{\alpha}}^{\textrm{ee}}$, $\bar{\bar{\alpha}}^{\textrm{mm}}$, $\bar{\bar{\alpha}}^{\textrm{me}}$, and $\bar{\bar{\alpha}}^{\textrm{em}}$  are electric, magnetic, magnetoelectric, and electromagnetic \textit{collective polarizability} tensors, respectively. The last two tensors $\bar{\bar{\alpha}}^{\textrm{em}}$ and $\bar{\bar{\alpha}}^{\textrm{me}}$ are referred to as bianisotropic ones~\cite{biama}. Next, assuming the time dependence ${\rm e}^{j\omega t}$ (where $\omega$ is the angular frequency), we write the incident electric field as ${\bf E}^{\rm i}=\left(E^{\rm i}_{x}\hat{{\bf x}}+E^{\rm i}_{y}\hat{{\bf y}}\right) e^{\pm jk z}$ propagating in $\mp z$ direction, where $k$ is the propagation constant in free space, and $E^{\rm i}_{x}$ and $E^{\rm i}_{y}$ are the $x$ and $y$ components of the incident field amplitude coefficients (with equal phases), respectively. We define the relation between the transmitted electric fields using the transmission dyadic (a tensor of second rank) $\bar{\bar{t}}$ as ${\bf E}^{\rm t}=\bar{\bar{t}}\cdot {\bf E}^{\rm i}$,  which in Cartesian coordinates reads 
\begin{equation}\label{T_LPbasis}
         \left[\begin{array}{c}
                 E^{\rm t}_{x} \\
                 E^{\rm t}_{y} \\
               \end{array}
             \right] = \left[
               \begin{array}{cc}
                 t_{xx} & t_{xy} \\
                 t_{yx} & t_{yy} \\
               \end{array}
             \right]\left[
               \begin{array}{c}
                 E^{\rm i}_{x} \\
                 E^{\rm i}_{y} \\
               \end{array}
             \right].
\end{equation}
Here, superscripts ``${\rm i}$'' and ``${\rm t}$'' correspond to the incident and transmitted fields, respectively.
It can be shown that, by using boundary conditions, the co- ($xx$ and $yy$) and cross- ($xy$ and $yx$) components of the transmission coefficients read (see Ref.~\cite{Niemi} for more details) \begin{equation}\label{Txx}
t^{\pm}_{xx}=1-{j\omega\over{2S}} \left(\eta{\alpha}^{\textrm{ee}}_{xx}+{{\alpha}^{\textrm{mm}}_{yy}\over \eta}\right),
\end{equation}
\begin{equation}\label{Tyy}
t^{\pm}_{yy}=1-{j\omega\over{2S}} \left(\eta{\alpha}^{\textrm{ee}}_{yy}+{{\alpha}^{\textrm{mm}}_{xx}\over \eta}\right),
\end{equation} \begin{equation}\label{Txyyx}
t^{\pm}_{xy}=t^{\mp}_{yx}=-{j\omega\over{2S}} \left[{(\eta{\alpha}^{\textrm{ee}}_{xy}-{{\alpha}^{\textrm{mm}}_{yx}\over \eta})}{\pm({\alpha}^{\textrm{em}}_{xx}+{\alpha}^{\textrm{em}}_{yy})}\right],
\end{equation}
respectively. Here ``$^+$'' and ``$^-$'' superscripts, respectively, correspond to the illumination from top and bottom [see Fig.~\ref{fig:fig1}], and we applied the reciprocity relations $
\bar{\bar{{\alpha}}}^{\textrm{ee}}=\left({\bar{\bar{\alpha}}}^{\textrm{ee}}\right)^{T}
$, $\bar{\bar{{\alpha}}}^{\textrm{mm}}=\left({\bar{\bar{\alpha}}}^{\textrm{mm}}\right)^{T}$, and $\bar{\bar{{\alpha}}}^{\textrm{em}}=-\left({\bar{\bar{\alpha}}}^{\textrm{me}}\right)^{T}$ [superscript ``${ T}$'' denotes the transpose operator] for polarizabilities. Moreover, $\eta$ is the characteristic wave impedance in the host medium. The reflection coefficients are obtained similarly and read \begin{equation}\label{Rxx}
r^{\pm}_{xx}=-{j\omega\over{2S}} \left(\eta{\alpha}^{\textrm{ee}}_{xx}-{{\alpha}^{\textrm{mm}}_{yy}\over \eta} {\mp 2 {\alpha}^{\textrm{em}}_{xy}}\right),
\end{equation}
\begin{equation}\label{Ryy}
r^{\pm}_{yy}=-{j\omega\over{2S}} \left(\eta{\alpha}^{\textrm{ee}}_{yy}-{{\alpha}^{\textrm{mm}}_{xx}\over \eta} {\pm 2 {\alpha}^{\textrm{em}}_{yx}}\right),
\end{equation} \begin{equation}\label{Rxyyx}
r^{\pm}_{xy}=r^{\pm}_{yx}=-{j\omega\over{2S}} \left[{(\eta{\alpha}^{\textrm{ee}}_{xy}+{{\alpha}^{\textrm{mm}}_{yx}\over \eta})}{\pm({\alpha}^{\textrm{em}}_{xx}-{\alpha}^{\textrm{em}}_{yy})}\right].
\end{equation}
Based on a general classification presented in Ref.~\cite{biama}, polarizability components ${\alpha}^{\textrm{em}}_{xx,yy}$ represent {\it chiral} properties in the metasurface whereas polarizability components ${\alpha}^{\textrm{em}}_{xy,yx}$ describe {\it omega} properties of the metasurface. There are several important observations from Eqs.~(\ref{Txx})--(\ref{Rxyyx}) which are summarized below (see also Ref.~\cite{Asadchy2}).

\emph{observation I}: For normal illumination, only the transverse components of the polarizability dyadics contribute to the transmission and reflection  coefficients, and the contributions of the normal to the metasurface plane components are absent.

\emph{observation II}: Co-component electric and magnetic polarizabilities contribute to the co-component transmission and  reflection coefficients whereas cross-component magnetoelectric polarizabilities ${\alpha}^{\textrm{em}}_{xy,yx}$ which represent the omega properties contribute to the co-component reflection coefficients only. More importantly, an asymmetric behavior in the co-component reflection from opposite directions is possible only when the metasurface possesses omega properties, i.e., $r_{xx}^+ \neq r_{xx}^-$ and/or $r_{yy}^+ \neq r_{yy}^-$ only if ${\alpha}^{\textrm{em}}_{xy}\neq 0$ and/or ${\alpha}^{\textrm{em}}_{yx}\neq 0$. 

\emph{observation III}: Cross-component electric and magnetic polarizabilities contribute to the cross-component transmission and reflection  coefficients whereas co-components magnetoelectric polarizabilities ${\alpha}^{\textrm{em}}_{xx,yy}$, which represent the chiral properties, contribute to the cross-component transmission and reflection coefficients only. More importantly, an asymmetric behavior in the cross-component transmission and reflection  coefficients from opposite directions is possible only when the metasurface exhibits chiral properties, i.e., $t_{xy}^+ \neq t_{xy}^-$ and/or $r_{xy}^+ \neq r_{xy}^-$ only if ${\alpha}^{\textrm{em}}_{xx}\neq 0$ and/or ${\alpha}^{\textrm{em}}_{yy}\neq 0$. 

\emph{observation IV}: There are two observations to identify if a metasurface certainly contains chiral property: to observe either asymmetric behavior in the cross-component transmission coefficients from opposite directions, i.e., $t_{xy}^+ \neq t_{xy}^-$ (or $t_{yx}^+ \neq t_{yx}^-$) or to observe asymmetric behavior between $xy$  and $yx$ components of the transmission coefficients from one side of the metasurface, i.e., $t_{xy}^{\pm} \neq t_{yx}^{\pm}$.

\emph{observation V}: A metasurface contains omega property only if one observes asymmetric behavior in the co-component reflection coefficients from opposite directions, i.e., $r_{xx}^+ \neq r_{xx}^-$ (or $r_{yy}^+ \neq r_{yy}^-$).

\subsection{Illumination with CP Plane Waves}
Although the analysis of a metasurface under the illumination by an LP plane wave provides all necessary information, the analysis of metasurfaces under the illumination by a CP plane wave provides important information specially for the case of metasurfaces with chiral properties. By considering an incident electric field as ${\bf E}_{\rm R}^{\rm i}=E^{\rm i0}\left(\hat{{\bf x}} \pm j\hat{{\bf y}}\right){\rm e}^{\pm jk z}$ for the right-hand CP plane wave and ${\bf E}_{\rm L}^{\rm i}=E^{\rm i0}\left(\hat{{\bf x}} \mp j\hat{{\bf y}}\right){\rm e}^{\pm j k z}$ for the left-hand CP plane wave propagating in  the  $-z$ (upper sign) or $+z$ (lower sign) direction, we define the transmission matrix as \begin{equation}\label{T_CPbasis}
         \left[\begin{array}{c}
                 {\bf E}_{\rm R}^{\rm t} \\
                 {\bf E}_{\rm L}^{\rm t} \\
               \end{array}
             \right] =\left[
               \begin{array}{cc}
                 t_\textrm{RR} & t_\textrm{RL} \\
                 t_\textrm{LR} & t_\textrm{LL} \\
               \end{array}
             \right]\left[
               \begin{array}{c}
                 {\bf E}_{\rm R}^{\rm i} \\
                 {\bf E}_{\rm L}^{\rm i} \\
               \end{array}
             \right],
\end{equation}
in the CP basis. In this basis the complex unit vectors $\hat{\bf R}$ and $\hat{\bf L}$ as the right-hand and left-hand vectors are, respectively, defined as $\hat{\bf R}=\left(\hat{\bf x}\mp j\hat{\bf y}\right)/\sqrt{2}$ and $\hat{\bf L}=\left(\hat{\bf x}\pm j\hat{\bf y}\right)/\sqrt{2}$ in terms of the unit vectors in Cartesian coordinates. The top and bottom signs correspond to the illumination from top and bottom of the the metasurface in $-z$ and $+z$ directions, respectively [see Fig.~\ref{fig:fig1}]. It can be shown that the co- (i.e., coefficients with RR and LL subscripts) and cross- (i.e., coefficients with RL and LR subscripts) components of the transmission coefficients in terms of the polarizabilities are obtained as \begin{equation}
\begin{array}{l}
\displaystyle
t_{\rm RR}^{\pm}=1-{j\omega\over{4S}}\left[\eta \left({\alpha}^{\textrm{ee}}_{xx}+{\alpha}^{\textrm{ee}}_{yy}\right)+{1\over \eta} \left({\alpha}^{\textrm{mm}}_{xx}+{\alpha}^{\textrm{mm}}_{yy}\right)
\right.\vspace*{.2cm}\\
\displaystyle
\hspace*{2.7cm}\left.
{+ 2j \left({\alpha}^{\textrm{em}}_{xx}+{\alpha}^{\textrm{em}}_{yy}\right)}\right],
\end{array}
\label{tRR_CPbasis_app}
\end{equation} \begin{equation}
\begin{array}{l}
\displaystyle
t_{\rm LL}^{\pm}=1-{j\omega\over{4S}}\left[\eta \left({\alpha}^{\textrm{ee}}_{xx}+{\alpha}^{\textrm{ee}}_{yy}\right)+{1\over \eta} \left({\alpha}^{\textrm{mm}}_{xx}+{\alpha}^{\textrm{mm}}_{yy}\right)
\right.\vspace*{.2cm}\\
\displaystyle
\hspace*{2.7cm}\left.
{- 2j \left({\alpha}^{\textrm{em}}_{xx}+{\alpha}^{\textrm{em}}_{yy}\right)}\right],
\end{array}
\label{tLL_CPbasis_app}
\end{equation} \begin{equation}
\begin{array}{l}
\displaystyle
t_{\rm RL}^{\pm}=t_{\rm LR}^{\mp}=-{j\omega\over{4S}}\left[\eta \left({\alpha}^{\textrm{ee}}_{xx}-{\alpha}^{\textrm{ee}}_{yy}\right)-{1\over \eta} \left({\alpha}^{\textrm{mm}}_{xx}-{\alpha}^{\textrm{mm}}_{yy}\right) 
\right.\vspace*{.2cm}\\
\displaystyle
\hspace*{1.5cm}\left.
{\mp 2j \left(\eta {\alpha}^{\textrm{ee}}_{xy}-{{\alpha}^{\textrm{mm}}_{yx}\over \eta}\right)}\right].
\end{array}
\label{tRL_CPbasis_app}
\end{equation}
Similar relations for illumination from one side were obtained earlier in~\cite{cuesta}.
As seen from these equations, chirality is observed in the co-components as opposed to LP illumination which was observed in the cross-component transmission coefficients. Moreover, metasurface anisotropy is observed from the cross-component transmission coefficients and it is distinguished from chirality. Also, the reflection coeffcients in the CP basis read \begin{equation}
\begin{array}{l}
\displaystyle
r_{\rm RR}^{\pm}=-{j\omega\over{4S}}\left[ 
\eta \left({\alpha}^{\textrm{ee}}_{xx}-{\alpha}^{\textrm{ee}}_{yy}\right)+{1\over \eta} \left({\alpha}^{\textrm{mm}}_{xx}-{\alpha}^{\textrm{mm}}_{yy}\right)
\right.\vspace*{.2cm}\\
\displaystyle
\hspace*{0cm}\left.
{\mp 2 \left({\alpha}^{\textrm{em}}_{xy}+{\alpha}^{\textrm{em}}_{yx}\right)}{\pm 2j \left(\eta{\alpha}^{\textrm{ee}}_{xy}+{{\alpha}^{\textrm{mm}}_{yx}\over \eta}\right)} {+2j \left({\alpha}^{\textrm{em}}_{xx}-{\alpha}^{\textrm{em}}_{yy}\right)}
\right],
\end{array}
\label{RR_CPbasis_app}
\end{equation}
\begin{equation}
\begin{array}{l}
\displaystyle
r_{\rm LL}^{\pm}=-{j\omega\over{4S}}\left[ 
\eta \left({\alpha}^{\textrm{ee}}_{xx}-{\alpha}^{\textrm{ee}}_{yy}\right)+{1\over \eta} \left({\alpha}^{\textrm{mm}}_{xx}-{\alpha}^{\textrm{mm}}_{yy}\right)
\right.\vspace*{.2cm}\\
\displaystyle
\hspace*{0cm}\left.
{\mp 2 \left({\alpha}^{\textrm{em}}_{xy}+{\alpha}^{\textrm{em}}_{yx}\right)}{\mp 2j \left(\eta{\alpha}^{\textrm{ee}}_{xy}+{{\alpha}^{\textrm{mm}}_{yx}\over \eta}\right)}{- 2j \left({\alpha}^{\textrm{em}}_{xx}-{\alpha}^{\textrm{em}}_{yy}\right)}
\right],
\end{array}
\label{LL_CPbasis_app}
\end{equation}
\begin{equation}
\begin{array}{l}
\displaystyle
r_{\rm RL}^{\pm}=r_{\rm LR}^{\pm}=-{j\omega\over{4S}}\left[\eta \left({\alpha}^{\textrm{ee}}_{xx}+{\alpha}^{\textrm{ee}}_{yy}\right)-{1\over \eta} \left({\alpha}^{\textrm{mm}}_{xx}+{\alpha}^{\textrm{mm}}_{yy}\right) 
\right.\vspace*{.2cm}\\
\displaystyle
\hspace*{2.6cm}\left.
{\mp 2 \left({\alpha}^{\textrm{em}}_{xy}-{\alpha}^{\textrm{em}}_{yx}\right)}\right].
\end{array}
\label{RL_CPbasis_app}
\end{equation}
Indeed, similar observations as in the case of an LP plane wave illumination are concluded for a CP plane wave illumination. The metasurface omega property can be observed from cross-component reflection coefficients. In general and in a loose language, the role between the co- and cross-component transmission and  reflection  coefficients in the LP and CP bases are interchanged in the analysis of metasurfaces.

\subsection{Applications and  Potentials}
Polarization transformation is one of the major applications of bianisotropic metasurfaces which is obtained by using chiral effects. For instance, a circular-polarization-selective metasurface with chiral property is able to reflect (or transmit) an incident wave  with only one handedness, i.e, right or left~\cite{Roy}. Also, when an LP incident wave passes though a properly engineered chiral metasurface, its polarization can be rotated. This phenomenon is called optical activity. In order to design an optically active metasurface insensitive to the polarization of   incident waves, one must compose it of chiral inclusions which are uniformity distributed in the metasurface plane~\cite{Niemi}. Another important application which is possible only using bianisotropic metasurfaces (notice that our focus here is on reciprocal metasurfaces) is asymmetric reflection (and also absorption) in co-polarization. This is possible using a metasurface with omega property~\cite{Yazdi}. Other applications such as   zero forward and backward scattering~\cite{Karilainen}, total absorption~\cite{Raadi} and one-way transparency~\cite{Raadi1}  are obtainable using bianisotropic metasurfaces, which will be discussed during the presentation.

\section{Gradient Bianisotropic Metasurfaces}
Although uniform metasurfaces provide many interesting applications, there are much more  functionalities if the uniformity condition is relaxed. The examples include but are not limited to anomalous refraction and/or reflection~\cite{Asadchy,Estakhri}, focusing acompanied by polarization manipulation~\cite{Veysi}, generation of structured vortex electromagnetic beams~\cite{Veysi2}, and others. There has been a fast growing metasurface category that is composed of inclusions supporting linear phase gradients~\cite{Asadchy2}. Indeed, it is shown in Refs.~\cite{Asadchy,Wong} that for the design of perfect-refraction gradient metasurfaces, the omega-type bianisotropy is required. Moreover, as we will show during the presentation, a perfect anomalous refraction together with an ideal polarization rotation is possible using chiral gradient metasurfaces~\cite{Kazemi}. We will review the current literature about different applications and potentials of metasurfaces   and give some possible ideas towards future road-map of this interesting research direction.

\section{Conclusions}
We reviewed the expressions for reflection and transmission coefficients in terms of the characteristic parameters of a bianisotropic metasurface when is excited normally by a plane wave with linear or circular polarization from either forward or backward direction. We distinguished between illuminations from opposite directions and provided interesting observations from the derived formulations both in the linear and circular polarization bases. We discussed two different asymmetry effects occurring in reflection due to the omega property and in transmission due to the chiral property of the metasurface (for  the linear basis). We have also briefly discussed some observations in the circular basis and potential applications of uniform and gradient metasurfaces.

\section{Acknowledgements}
This work was supported by the W. M. Keck Foundation (USA).

\end{document}